\documentclass[12pt]{article}
\usepackage{psfig,epsf}
\usepackage{a4wide,amssymb,graphicx}
\usepackage{cite}

\begin{document}

\title{Helicity asymmetries in double pion photoproduction on the
proton}

\maketitle

\begin{center}
{\large{L.~Roca}}

\vspace{.5cm}
{\it Departamento de F\'{\i}sica Te\'orica and IFIC,
 Centro Mixto Universidad de Valencia-CSIC\\
 Institutos de Investigaci\'on de Paterna, Apdo. correos 22085,
 46071, Valencia, Spain}

\end{center}

 \begin{abstract} 

Based on a prior model on double pion photoproduction on the 
proton, successfully tested in total cross sections and invariant
mass distributions, we make a theoretical study of the angular
dependence of helicity asymmetries from the interaction of
circularly polarized photons with unpolarized protons. 
We show that this observable is sensitive to details of the
internal mechanisms and, thus, represents a complementary test
of the theoretical model.

\end{abstract}

\section{Introduction} 
The photoproduction of two pions on nucleons at low and
intermediate energies (up to $E_\gamma\sim 1\textrm{ GeV}$) has
been the subject of intense experimental 
\cite{Braghieri:1994rf,Zabrodin:1997xd,Harter:jq,Wolf:qt,Assafiri:mv,Kotulla:2003cx}
and theoretical 
\cite{luke,Murphy:1996ms,tejedor1,tejedor2,
roberts,Ochi:1997ev,Hirata:1997gx,
Ripani:2000va,roca,Bernard:2001gz,Ong:2001bm,Hirata:2002tp,fix}
 study.
 The works have been mainly motivated
to understand the role of the many baryonic resonances involved
in the process. The fact that there are three particles in the
final state, $N\pi\pi$, gives way to different mechanisms in
which baryonic resonances play an important role, and this has
led to obtain useful information on some resonances not
attainable with other reactions.
Much of the work has been done in unpolarized observables,
mostly total cross sections and invariant mass distributions. 
In particular, the work of \cite{roca}, which is the one we will
use along this work, is based on around $25$ Feynman diagrams
considering the coupling of photons to several baryonic
resonances able to influence the energy region up to
$E_\gamma\sim 800\textrm{ MeV}$. The main advantage is the use
of no free parameters. This model succeeded in reproducing total
cross sections and invariant mass distributions for all the
charge channels with a good accuracy, not only in
nucleons but also in nuclei. Specially remarkable was the
application of the model of \cite{roca} to the study of the
photoproduction of two pions in nuclei \cite{rocasigma}. It
succeeded in describing the shift of strength in the double
pion  invariant mass distribution towards the $2m_\pi$ masses,
due to the modification of the $\sigma$-meson mass in nuclear
matter, where the $\sigma$-meson is dynamically generated as a
$\pi\pi$ rescattering in scalar-isoscalar channel. This
prediction was confirmed by the experiment of \cite{metag}.
Therefore, the model of \cite{roca} has widely proven his
efficiency in reproducing and predicting unpolarized observables
in the energy range from threshold up to  $E_\gamma\sim
800\textrm{ MeV}$. Nonetheless, a more demanding test to the
model can be done by evaluating polarization observables, since
it can be sensible to details of the model not visible when
integrating over polarization degrees of freedom in the
unpolarized observables.  In this line, a test of the model of
\cite{roca} was done in \cite{nacherpol} when evaluating
 the spin $1/2$ and $3/2$ amplitudes and the
contribution to the GDH sum rule of the double pion channel, in
fair agreement with Mainz results \cite{Lang:2002fj,Ahrens:2003na},
and the
evaluation of  beam asymmetries under experimental study at
GRAAL \cite{Assafiri:mv}. These observables are based on
differences of total differential cross sections dependent on
polarization, which 
provide a valuable information on the internal dynamics of
the reaction. However, these observables still rely on integrated
cross sections and no angular distributions are provided from
where more information can be obtained.

The aim of the present work is to evaluate
 angular dependences of the
cross section asymmetry $\sigma^+-\sigma^-$  for the absorption
of circularly polarized photons by unpolarized protons.
 This
observable is very sensitive to the internal mechanisms of the
reaction and, therefore, can be a very useful test to impose
constraints on the theoretical models. The work has been partly
motivated by preliminary experimental results, for the 
$\gamma p\to\pi^+\pi^- p$ channel, with the CLAS
detector at Jefferson Lab \cite{strauch} which shows strong and
not trivial angular dependences of this observable, and prospects
of measurements at Mainz \cite{beck} for the 
$\gamma p\to\pi^0\pi^0 p$ channel.


\section{Summary of the $\gamma p\to\pi\pi p$ model}

\begin{figure}
\centerline{\protect\hbox{
\psfig{file=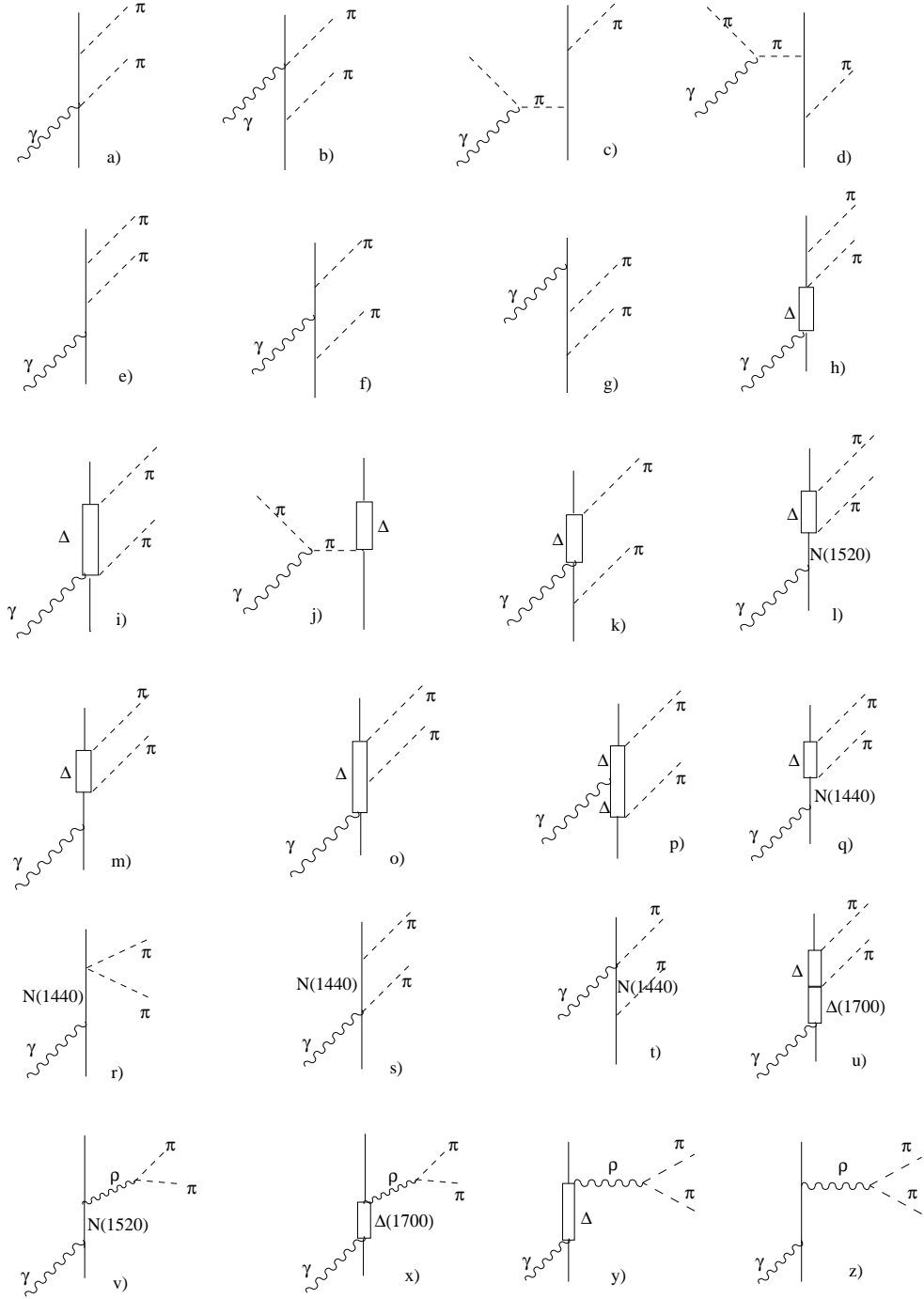,width=0.8\textwidth,silent=}}}
\caption{Mechanisms used in the model for $\gamma
p\rightarrow\pi^+\pi^- p$. Solid lines without labels are nucleons.
$\Delta$ means $\Delta(1232)$. For the 
$\gamma p\to\pi^0\pi^0 p$ channel only the $e$, $f$, $g$,
 $h$, $k$, $l$, $m$, $o$, $p$, $q$, $r$ and $u$ mechanisms
contribute.}
\label{fig:diagrams}
\end{figure}

In this section we briefly summarize the model of
\cite{tejedor1,tejedor2,roca} for the double pion
photoproduction on nucleons. This model is intended to reproduce
the total cross sections and invariant mass distributions up to
photon energies of $E_\gamma\sim 800 \textrm{ MeV}$.  The
model is based on a set of tree level mechanisms, depicted in
Fig.~\ref{fig:diagrams}, for the $\pi^+\pi^-$ channel. 
For the  $\pi^0\pi^0$ channel only the mechanisms $e$, $f$, $g$,
 $h$, $k$, $l$, $m$, $o$, $p$, $q$, $r$ and $u$ contribute.
 These Feynman diagrams involve pions, $\rho$-mesons, nucleons
 and nucleonic and $\Delta$ resonances.
 The baryon resonances included
in the model are: $\Delta$(1232) or $P_{33}$ ($J^{\pi}=3/2^+$, I=3/2), 
$N^\ast$(1440) or $P_{11}$ ($J^{\pi}=1/2^+$, I=1/2), 
$N^\ast$(1520) or $D_{13}$ ($J^{\pi}=3/2^-$, I=1/2) and
$\Delta$(1700) or $D_{33}$ ($J^{\pi}=3/2^-$, I=3/2). The
contribution of the $N^\ast(1440)$  is small but it was included
due to the important role played by that resonance in the $\pi
N\rightarrow\pi\pi N$ reaction and the fact that the excitation
of the $N^\ast(1440)$ peaks around $600 \textrm{ MeV}$ photon
energy in the $\gamma N$ scattering. The $N^\ast(1520)$ has a
large coupling to the photons and is an important ingredient due
to its interference with the dominant term of the process, the
$\gamma N\rightarrow\Delta\pi$ transition called the 
$\Delta$-Kroll-Ruderman ($\Delta$KR) contact term. 
(The $\Delta$KR term is not present in the 
$\gamma p\to\pi^0\pi^0 p$ channel).
Several $\rho$ and $\Delta$(1700)
terms were included in the last version of the model \cite{roca}
because of important interference effects. The consideration
 of the
$\rho$ terms was of crucial importance in the analysis of
Ref.~\cite{rocarho}, when studying $\rho$ meson photoproduction
in nuclei. The $y)$ and $z)$ diagrams considering a $\rho$
exchange were not considered in \cite{roca} since they give
negligible contribution to the cross section in the energy region
of concern. However they were
considered in the work of \cite{rocarho} by completeness when
considering $\rho$ meson photoproduction and we also include them
here since they can produce a non-negligible influence in the
polarization asymmetry.

 No other resonances were considered in the model since they
  cannot
appreciably change the results in the energy range up to
$E_\gamma\sim 800 \textrm{ MeV}$, because their widths are small and
lie at too high energies, because the helicity amplitudes are
small, because the decay width rates into $\Delta\pi$ or $\rho
N$ are small or because a combination of various of these
effects \cite{roca}.

 The diagrams u), v) and x) of Fig.~\ref{fig:diagrams}  are
the main modifications of  \cite{roca} with respect to
\cite{tejedor2}. In the first work  of \cite{tejedor1} they
included more than $50$ diagrams for the 
$\gamma p\to\pi^+\pi^-p$ channel, but many of them were shown to be
negligible at energies up to $E_\gamma=800 \textrm{ MeV}$. The
non-negligible contributions come from  the diagrams of
Fig.~\ref{fig:diagrams}.

The amplitudes are evaluated from effective interaction
Lagrangians which are shown in the Appendices of
Ref.~\cite{roca}, using a non relativistic approximation exact
up to order $p/M_p$, that is, removing terms of order
$(p/M_p)^2$ and higher.

It is important to stress that this model has no free parameters,
in the sense that there is no parameter to be fitted to the
experimental double pion photoproduction observable. All input
needed is obtained uniquely from properties of resonances and
their decays. Where there are doubts about relative signs of
couplings, one resorts to quark models or chiral perturbation
theory to fix them \cite{cano}.

\section{Photon helicity asymmetry}

We will consider the absorption of circularly polarized
photons by non-polarized protons. 
For real photons the polarization
vectors of a circularly polarized photon can be expressed as:
\begin{equation}
\vec{\epsilon}\,^\pm =\frac{1}{\sqrt{2}}(\mp 1,-i,0)
\label{eq:epsilon}
\end{equation} 
where $\vec{\epsilon}\,^+$ or $\vec{\epsilon}\,^-$
represent a right-handed or left-handed circularly
polarized photon respectively.

The helicity asymmetry that we are going to consider in the
present work can be defined as 
\begin{equation}
A\equiv\frac{d\sigma^+-d\sigma^-}{d\sigma^++d\sigma^-}
\label{eq:A}
\end{equation}
where $d\sigma^{+(-)}$ is the differential cross section for the
interaction of a right-handed (left-handed) circularly polarized
photon with an unpolarized proton.

\begin{figure}
\centerline{\protect\hbox{
\psfig{file=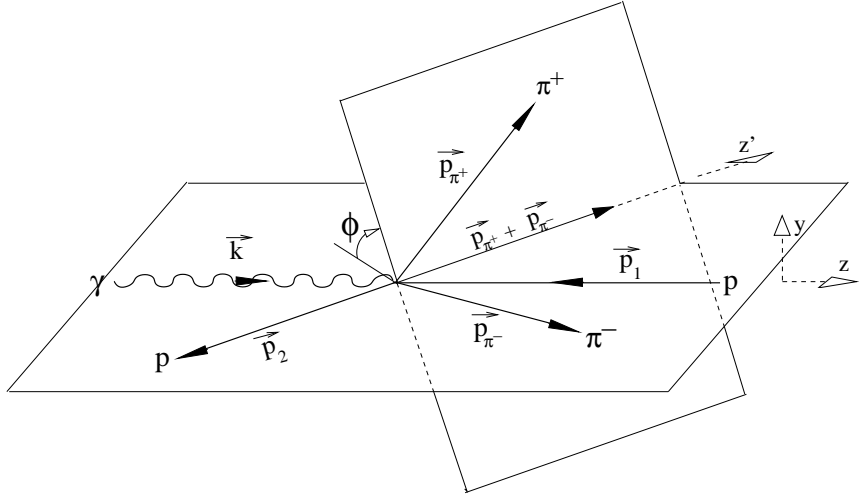,width=0.7\textwidth,silent=}}}
\caption{Angular and kinematics definition in the
helicity frame. The angle $\phi$ represents the angle between the
scattering plane ($\vec k\vec p_2$) and the two pions plane 
($\vec p_{\pi^+}\vec p_{\pi^-}$). (See the text for the exact definition
for the origin and sign of $\phi$).}
\label{fig:kinematics}
\end{figure}

Three different frames have been commonly used in the literature
to describe angular distributions in three body final state
processes \cite{ABBHHM,Schilling:1969um,Ballam:1971yd}. These
frames are called {\it Gottfried-Jackson}, {\it helicity}
 and {\it Adair} systems,
differing in the choice of the $z'$ axis from which the
azimuthal and polar angles are defined.
The choice of a
particular frame is of relevance when studying particular
production processes, like vector-meson photoproduction, due to
particular angular distributions for a certain spin of the
intermediate meson. In the present work, in order to test 
all the possible mechanisms contributing to
the  $\gamma p\to\pi^+\pi^- p$ process, any of these frames is
useful for our purposes. Therefore, in order to allow comparison
with preliminary experimental results with the CLAS detector at
Jefferson Lab \cite{strauch},
 we will use the {\it helicity}
frame, which is defined as having the $z'$ axis in the direction
of the sum of the momentum of both pions in the overall ($\gamma
p$) c.m. frame. The use of the other frames would produce
similar
qualitative results for the discussion done in the present work.
In Fig.~\ref{fig:kinematics} the kinematics for the 
$\gamma (k)p(p_1)\to\pi^+(p_{\pi^+})\pi^-(p_{\pi^-})p(p_2)$
reaction in this {\it helicity} frame is shown.

The $\phi$ angle, which we will use along the present work,
accounts for the angle between the scattering plane (containing
the photon, $\vec k$, and the final proton, $\vec p_2$) and the
plane containing the two pions, and is defined by:

\begin{eqnarray}
\cos \phi &=& \frac{(\vec k\times\vec q)\cdot (\vec
q\times\vec p_{\pi^+})}
{|\vec k\times\vec q||\vec
q\times\vec p_{\pi^+}|} \nonumber \\
\sin \phi &=& -\frac{((\vec k\times\vec q)\times \vec q)\cdot (\vec
q\times\vec p_{\pi^+})}
{|\vec (k\times\vec q)\times \vec q||\vec q\times\vec p_{\pi^+}|}
\label{eq:defphi}
\end{eqnarray}
with $\vec q\equiv \vec p_{\pi^+}+\vec p_{\pi^-}$, and with all the
momenta in the overall center of mass frame.
The expressions in  Eq.~(\ref{eq:defphi}) establish the definition
of $\phi$ without ambiguities in the origin or sign, and uses the
same convention as in Ref.~\cite{Schilling:1969um}.

In the present work we will consider the $\phi$ dependence of
the $A$ observable since it is very sensitive to the particular
mechanisms involved in the reaction by different reasons. First,
it is sensible to differences of cross sections,
($\sigma^+-\sigma^-$), and thus
mechanisms which give small contribution to the total cross
section can produce a sizeable contribution to the helicity
asymmetry if the mechanisms are strongly helicity dependent.
Second, it is very sensitive to interferences between the different
diagrams of the model, as we explain in detail below.

Let us write  the amplitude for the process as
\begin{equation}
T=\epsilon_\mu T^\mu.
\end{equation}
In order to evaluate the cross sections one has to consider the
squared $T$-matrix averaged over the initial spin of the 
proton, since
we are considering  non-polarized target,
and summed over the spins of the final proton:
\begin{equation} 
\sum_{s_i,s_f} \langle m_{s_i}|\epsilon_\mu T^\mu
|m_{s_f}\rangle \langle m_{s_f} |\epsilon^*_\nu T^{\dagger\nu}
|m_{s_i}\rangle
=Tr\{\epsilon_\mu T^\mu \epsilon^*_\nu T^{\dagger\nu}\}.
\label{eq:trace}
\end{equation}

By using Coulomb gauge, where $\epsilon_0=0$ and transversality,
$\epsilon_z=0$, with $\hat z$ the direction of the photon
momentum ($\vec{k}$), Eq.~(\ref{eq:trace}) reads
\begin{equation}
Tr\{|\epsilon_x|^2T_xT_x^\dagger+|\epsilon_y|^2T_yT_y^\dagger
+\epsilon_x\epsilon^*_yT_xT_y^\dagger
+\epsilon^*_x\epsilon_yT_yT_x^\dagger\}.
\end{equation}

Should we use linearly polarized photons
($\vec{\epsilon}=(1,0,0)$ or $(0,1,0)$),
we would obtain for the numerator of Eq.(\ref{eq:A}) 
up to phase space integrals,
$Tr\{T_xT_x^\dagger-T_yT_y^\dagger\}$
\footnote{In the nomenclature of "hadronic tensor",$W^{\mu\nu}$,
and "structure functions" it would be
$\sim(W^{xx}-W^{yy})=W_{TT}$. (See, for instance,
ref.~\cite{boffi}).}
 (which is roughly the 
{\it beam asymmetry} $\Sigma$ already studied in
\cite{nacherpol}).
If we use circularly polarized photons, the numerator of
Eq.(\ref{eq:A}) goes as (up to phase space integrals):

\begin{equation}
d\sigma^+-d\sigma^-\sim -2Tr\{i(T_xT_y^\dagger-T_yT_x^\dagger)\}
=4{\cal I}m\{Tr\{T_xT_y^\dagger\}\}=
-2i Tr\{(\vec T \times \vec{T^\dagger})_z\}
\label{eq:cross}
\end{equation}
where $(\vec T \times \vec{T^\dagger})_z$ means the component
in the photon direction
of the cross product of $\vec T$ and $\vec{T^\dagger}$.

There is an interesting necessary condition for the
helicity asymmetry $A(\phi)$, that is 
\begin{equation}
A(\phi)=-A(2\pi-\phi).
\label{eq:Aphi}
\end{equation}
This is true 
since the change $\phi\to (2\pi-\phi)$ can be interpreted as a
reflection of the $\vec p_{\pi^+}\vec p_{\pi^-}$ plane with
respect to the $\vec k\vec p_2$ plane (see
Fig.~\ref{fig:kinematics}). This is equivalent to changing the
sign of the $y$ coordinate and, therefore, by looking at 
Eq.~(\ref{eq:epsilon}), to the exchange of the role
right-handed$\leftrightarrow$left-handed, what means $A\to-A$.
The condition $A(\phi)=-A(2\pi-\phi)$ implies that $A$ can be
expanded as 
\begin{equation}
A(\phi)=\sum_{n=1}^\infty a_n\sin(n\phi), 
\textrm{ with } n=1,2,3,4,...
\label{eq:series}
\end{equation}

The proportionality on $\sin(\phi)$ implicit in
Eq.~(\ref{eq:series}), implies the asymmetry to be proportional
to $(\vec p_{\pi^-} \times \vec p_{\pi^+})_z$, since $\sin(\phi)$
is proportional to $(\vec p_{\pi^-} \times \vec p_{\pi^+})\cdot
\vec k$ as can be easily obtained from Eq.~(\ref{eq:defphi}).

This $(\vec p_{\pi^-} \times \vec p_{\pi^+})_z$ proportionality
relation can also be obtained from the general structure that
the double pion photoproduction amplitude can take, that was
shown in Ref.~\cite{roberts}. This general expression of the
amplitude can be obtained \cite{roberts} by using Lorentz
covariance and gauge invariance and can be written as:

\begin{eqnarray}
\vec{T}=F_1\vec p_{\pi^-}+F_2\vec p_{\pi^+}
+F_3 \vec\sigma\cdot(\vec p_{\pi^-}+\vec p_{\pi^+})
\vec\sigma\cdot\vec p_{\pi^-}\vec p_{\pi^-}
+F_4 \vec\sigma\cdot(\vec p_{\pi^-}+\vec p_{\pi^+})
\vec\sigma\cdot\vec p_{\pi^-}\vec p_{\pi^+} \nonumber\\
+F_5 \vec\sigma\cdot\vec p_{\pi^+}\vec\sigma\cdot\vec k 
\vec p_{\pi^-}
+F_6 \vec\sigma\cdot\vec p_{\pi^+}\vec\sigma\cdot\vec k 
\vec p_{\pi^+}
+F_7 \vec\sigma\cdot\vec p_{\pi^-}\vec\sigma\cdot\vec k 
\vec p_{\pi^-}
+F_8 \vec\sigma\cdot\vec p_{\pi^-}\vec\sigma\cdot\vec k 
\vec p_{\pi^+} \nonumber \\
+F_9 \vec\sigma\cdot\vec p_{\pi^-}\vec\sigma
+F_{10} \vec\sigma\cdot\vec p_{\pi^+}\vec\sigma
+F_{11} \vec k\times\vec \sigma
+F_{12} \vec\sigma\cdot(\vec p_{\pi^-}+\vec p_{\pi^+})
\vec\sigma\cdot\vec p_{\pi^-} (\vec k\times\vec \sigma).
\label{eq:generalT}
\end{eqnarray}

The $F_i$ coefficients are, in general, complex functions of
the momenta accounting for the dynamics
of the different mechanisms
(propagators, momentum dependences of the vertices, etc).
$\vec{T}$ has to change sign under parity (since
$\vec T\cdot\vec\epsilon$ has to be invariant and 
$\vec\epsilon$ is a
vector). This implies that all the $F_i$ coefficients are
scalars under parity. Time reversal symmetry implies also that
$F_i$, for $i=1-10$, are invariant under time reversal
but $F_{11}$ and
$F_{12}$ has to change.

After applying Eq.~(\ref{eq:cross}) to the amplitude in
Eq.~(\ref{eq:generalT}) we have checked that the resulting 
non-vanishing terms can always be rearranged in terms of the
form
\begin{equation}
d\sigma^+-d\sigma^-=\sum G\,{\cal I}m(\eta F_mF_n^*)
(\vec p_{\pi^-} \times \vec p_{\pi^+})_z,
\label{eq:cross2}
\end{equation}
Where the $G$'s are scalar functions of $\vec p_{\pi^-}$, 
$\vec p_{\pi^+}$ and $\vec k$.
In Eq.~(\ref{eq:cross2}), $\eta$ is $1$ if $F_mF_n^*$ is invariant
under time reversal or $i$ if it is not\footnote{Recall that time
reversal changes $i$ by $-i$.}.

Let us discuss some important consequences that can be concluded
from Eq.~(\ref{eq:cross2}) on the sensitivity of the helicity
asymmetry to the internal structure of the mechanisms and to the
interferences between different diagrams.  In some 
mechanisms\footnote{In the following discussion we will call
"mechanism" to the different graphs of Fig.~\ref{fig:diagrams}
without specifying the charge configuration, and we will call
"diagram" to the different charge configurations that a mechanism
can have.} it
is allowed to have either a $\pi^+$ or a $\pi^-$ in both the two
external pion lines (we will call it type-I mechanisms), while in
other mechanisms only one charge configuration is possible
(type-II). (Type-II mechanisms are $a$, $b$, $c$, $d$, $e$, $f$,
$g$, $h$, $k$, $p$, $s$, $t$, $v$, $x$, $y$ and $z$ of
Fig.~\ref{fig:diagrams} and type-I are the rest). For instance,
in the $\Delta$KR term ($i$ mechanism of Fig.~\ref{fig:diagrams})
it is possible to have a $\pi^+$ in the $\gamma N \Delta\pi$
vertex and a $\pi^-$ in the $\Delta N\pi$ vertex or vice-versa,
while in the $a$ mechanism only the diagram where
the $\pi^+$ is in the $\gamma NN\pi$ vertex and the $\pi^-$ in the
$NN\pi$ vertex is possible.
If we take any individual diagram the asymmetry will
necessarily vanish. 
This is so since the general complex structure of propagators is
the same in all the pieces of the amplitude of 
Eq.~(\ref{eq:generalT}) if one considers only one diagram,
and therefore the $F_i$ coefficients can be factorized as
$F_i=\alpha a_i$, where $\alpha$ contains all the structure of
propagators and $a_i$ are real for $i=1-10$ and purely imaginary
for $i=11,12$. Therefore, in Eq.~(\ref{eq:cross2}) we have
$|\alpha|^2{\cal I}m(\eta a_ma_n^*)$ with $\eta a_ma_n^*$ being
real. On the other hand, if we have two diagrams  then the
$\alpha$ coefficient can be different for some terms of the
amplitude and the $|\alpha|^2$ factorization does not hold.
In conclusion, associating mechanisms to diagrams, only the
mechanisms that have two possible diagrams associated, i.e.,
those of type-I, could by themselves be nonzero
(although some of them can also be zero).
 On the other
hand, the fact that one needs ${\cal
I}m(\eta F_m F_n^*)\ne 0$ implies the coefficients $F_i$ 
to be generally complex, and this is provided by the 
propagator structure of the
diagrams. This is the reason why this observable is so sensitive
to the internal mechanisms of the reaction. In addition,
the fact that the $(\vec p_{\pi^-} \times \vec p_{\pi^+})_z$ 
factor of
Eq.~(\ref{eq:cross2}) makes the numerator  of $A$ to be
proportional to $\sin \phi$ implies that the difference from a
simple $\sin \phi$ dependence comes from the momentum dependence
of the $F_i$ coefficients of the amplitude. Thus, the
angular dependence of the helicity asymmetry is strongly
reflecting the internal structure of the various mechanisms.

On the other hand, when allowing the different mechanisms to
interfere between them, even if some individual mechanisms
do not produce
an asymmetry by themselves, they can produce a non-vanishing
asymmetry when adding them coherently. This is why the
interferences in the $\phi$ dependence of the helicity asymmetry
are so important. The angular dependence of the denominator of
the helicity asymmetry, $d\sigma^++d\sigma^-$ (which is
proportional to the total cross section), does not modify
qualitatively the previous discussion.

Let us illustrate the previous discussion with an example:
let us consider the $\Delta$KR term ($i$ mechanism of
Fig.~\ref{fig:diagrams}). The amplitude for the process where the
$\pi^+$ is emitted before the $\pi^-$ is given by \cite{roca}
\begin{equation}
\vec{T}_{\Delta KR} = \frac{1}{9} e(\frac{f^\ast}{m_\pi})^2
G_\Delta(p_2 + p_{\pi^-}) F_\pi((p_{\pi^+}-k)^2) [
2\vec{p}_{\pi^-} - i (\vec{\sigma}\times\vec{p}_{\pi^-}) ],
\end{equation}
where $f^\ast=2.13$, $G_\Delta$ is the $\Delta(1232)$ propagator
and $F_\pi$ is a form factor. For the process where
the $\pi^-$ is emitted before the $\pi^+$ the amplitude is
obtained by exchanging  $p_{\pi^+}\leftrightarrow p_{\pi^-}$ and
writing the appropriate isospin coefficients
\begin{equation}
\vec{T}_{\Delta KR} = -\frac{1}{3} e(\frac{f^\ast}{m_\pi})^2
G_\Delta(p_2 + p_{\pi^+}) F_\pi((p_{\pi^-}-k)^2) [
2\vec{p}_{\pi^+} - i (\vec{\sigma}\times\vec{p}_{\pi^+}) ].
\end{equation}

With these expressions for the amplitude,
we have for the last term of Eq.~(\ref{eq:cross})
\begin{equation}
Tr\{(\vec T \times \vec{T^\dagger})_z\}=20 i \,{\cal
I}m(\alpha\beta^*)(\vec p_{\pi^-} \times \vec p_{\pi^+})_z.
\end{equation}
with $\alpha=\frac{1}{9} e(\frac{f^\ast}{m_\pi})^2
G_\Delta(p_2 + p_{\pi^-}) F_\pi((p_{\pi^+}-k)^2)$ and 
$\beta=-\frac{1}{3} e(\frac{f^\ast}{m_\pi})^2
G_\Delta(p_2 + p_{\pi^+}) F_\pi((p_{\pi^-}-k)^2)$. This means
that the $\Delta$KR mechanism (accounting for two Feynman diagrams)
 gives by itself a non-vanishing angular 
dependence of the helicity asymmetry 
thanks to the imaginary part in the
$\Delta$ propagator, and deviation from a simple $\sin \phi$
dependence (as will be shown in the Results section) 
is due to the momentum dependence of the 
propagator and the form factor, (although this latter one is a
smooth function). 

The kind of reasoning presented in this section stresses the
importance of small details of the theoretical models, making
the $\phi$ dependence of the helicity asymmetry a very useful
and powerful tool to check the heart of the theoretical models.
Therefore, even if the model succeeds to reproduce unpolarized
observables (like total cross sections, invariant mass
distributions, etc), it could fail to reproduce the kind of
polarization observables studied in this work,
simply because of small details.

\section{Results}

\begin{figure}
\centerline{\protect\hbox{
\psfig{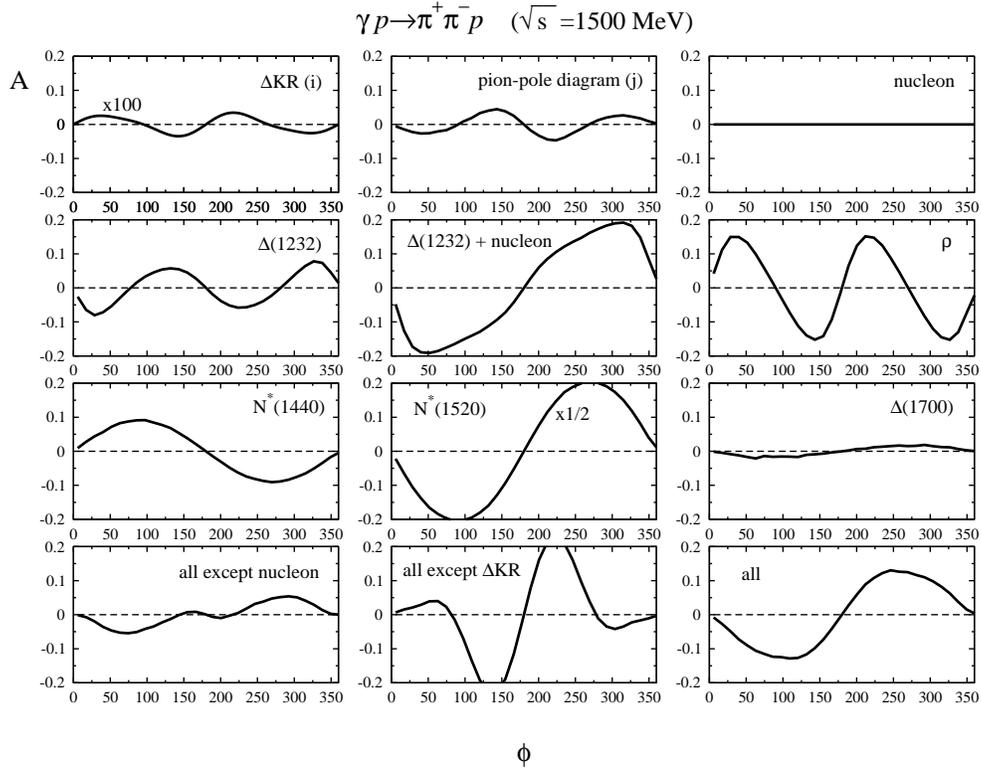}}}
\caption{Angular ($\phi$) distribution of the helicity asymmetry,
$A$,
for different contributions in the $\gamma p\to\pi^+\pi^- p$
channel for a $\gamma p$ energy of 
$\sqrt{s}=1500\textrm{ MeV}$.} 
\label{fig:results1}
\end{figure}

In Fig.~\ref{fig:results1} we show, for a given $\gamma p$
energy  of $\sqrt{s}=1500\textrm{ MeV}$  ($E_\gamma^{Lab}\simeq
730\textrm{ MeV}$), the $\phi$ distribution of the helicity
asymmetry, $A$, for different mechanisms contributing to the
$\gamma p\to\pi^+\pi^- p$
process. From left to right and up to down the plots
represent: $\Delta$KR term ($i$ diagram of
Fig.~\ref{fig:diagrams}), pion pole term ($j$ diagram), nucleon
intermediate mechanisms (diagrams from $a$ to $g$), 
$\Delta(1232)$ (diagrams $h$ to $k$ and $m$ to $p$),
$\rho$-meson intermediate contribution (diagrams $v$ to $z$),
$N^*(1440)$ resonance
 (diagrams $q$ to $t$), $N^*(1520)$ (diagrams
$l$ and $v$), $\Delta(1700)$ (diagrams
$u$ and $x$), all the
mechanisms except the nucleon intermediate diagrams, all the
mechanisms except the $\Delta$KR term and, finally, the full
model (all the diagrams).  (The plots of the $\Delta$KR term
alone and the $N^*(1520)$ have been multiplied by $100$ and $1/2$
respectively to make the curves visible inside the represented
scale). 
In the plots the condition $A(\phi)=-A(2\pi-\phi)$,
Eq.~(\ref{eq:Aphi}), is clearly visible.

One can see in Fig.~\ref{fig:results1}
 the very strong dependence on
the mechanisms considered and the crucial role of the
interferences. For instance, even if the nucleon intermediate
mechanisms give a vanishing contribution by themselves, the
interference with the $\Delta(1232)$ mechanisms produces
strong changes in the distribution with respect to considering
the $\Delta(1232)$ terms alone. 
From the discussion of the previous section it can be understood
why the nucleon mechanisms give a zero assymetry: the nucleon
propagators do not have width and therefore do not have a complex
structure needed to produce a non-zero imaginary part in
Eq.~(\ref{eq:cross2}). 
Another quantitative example of
the important role of the interferences can be seen, for
instance, by looking at the figures evaluated   with all the
mechanisms except the nucleons or except de $\Delta$KR term.
For this latter case, despite the asymmetry for the $\Delta$KR
being very small, it has an important influence in the full result.
On the other hand, by comparing the
 "all except nucleon" with the "all" plot, one
can see the dramatic influence of the nucleon intermediate
mechanism in the angular distribution of the helicity asymmetry,
despite these mechanisms contributing only around $10$\% to the
total cross section.

\begin{figure}
\centerline{\protect\hbox{
\psfig{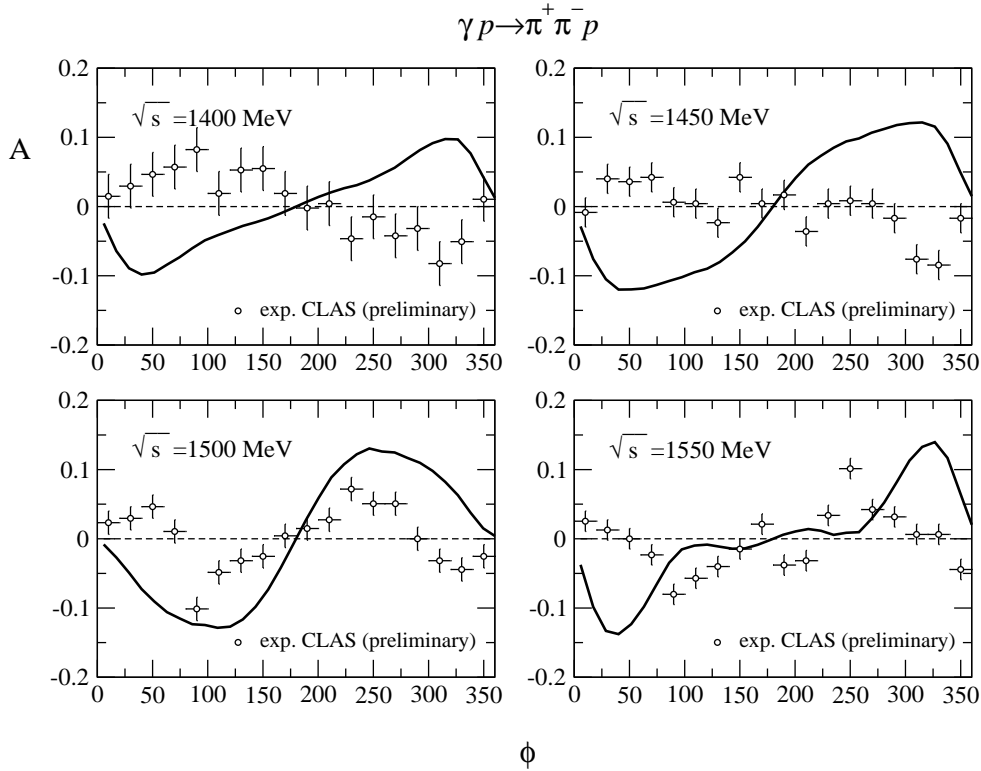}}}
\caption{Angular distribution of the helicity asymmetry for
different energies with the full model, for the
$\gamma p\to\pi^+\pi^- p$ channel. Preliminary experimental
results from \cite{strauch}. (The data is integrated over the
full CLAS acceptance while the theoretical calculations cover the
full phase space).} 
\label{fig:results2}
\end{figure}

In order to show the sensitivity to the energy of the  angular
distribution of the helicity asymmetry, we show in
Fig.~\ref{fig:results2} the results with the full model for
different energies. The experimental data, still preliminary,
are obtained from Ref.~\cite{strauch}, measured with the CLAS
detector at Jefferson Lab. It is important to stress that these
data are integrated over the full  CLAS acceptance,
while the theoretical
model covers the full phase space. Thus, given the sensitivity
of the observable to these details, one has to be cautious when
making conclusions from this naive comparison. With this caveat,
and after the remarks on the sensitivity of this observable to
small details of the model, the comparison of the theoretical
predictions of the present work and the data of \cite{strauch}
shown in Fig.~\ref{fig:results2} would be seen as an indication
that the model contains the basic mechanisms.
The strength of the theoretical results and experiment is
similar, and this is not a trivial theoretical result given the
large range of values found in Fig.~\ref{fig:results1} for
different options of partial results of the model. The
discrepancies found in the shape for the two lower energies are
more worrisome, but in view of the preliminary character of the
experimental data, and the fact that they are not $4\pi$
integrated, it is probably too early to draw conclusions from
there. We would like to note that the 
theoretical results reported here are
similar in strength and shape as those reported in \cite{strauch} as
private communication, calculated with the model of
\cite{mokeev}. In view of this, it is important that definitive
data are provided and that direct calculations adapted to the
acceptance of the experimental setup are carried out. This
comparison should help in the future to improve the present
models of double pion photoproduction.
We are aware, that given the important role of the sources of
complex part in the amplitudes in this polarized observable,
other sources of complex part not considered in our model could
be relevant for the polarization observables in spite that they
could be not so important in the unpolarized observables. For
instance, in Ref.~\cite{Strauch:2003rx} it was pointed out, as
private communication from Mokeev, that complex relative phases
in the amplitudes could produce sizeable effects in these
polarization observables, whereas in unpolarized observable they
do not.
 On the other hand, final state interaction of the produced
particles could also influence these observables. Therefore, we
are well aware of the limitations of our model, but the present
work can serve to establish, from the comparison with experiments,
how much room for improvement  one can
expect on the theoretical models. 
\\

\begin{figure}
\centerline{\protect\hbox{
\psfig{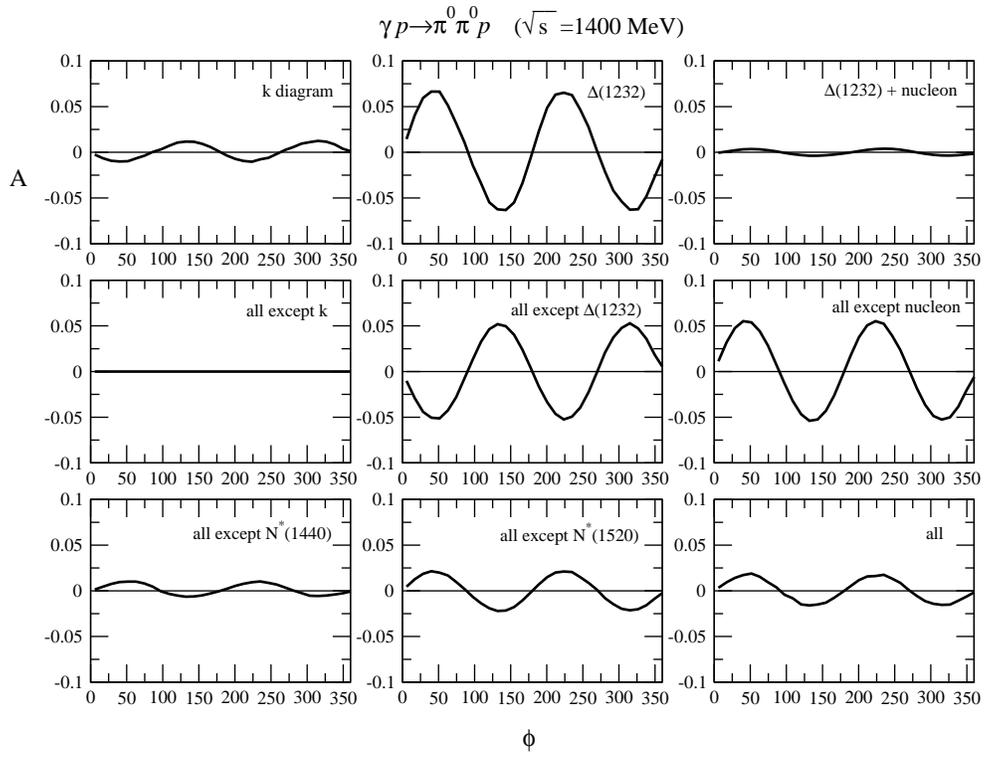}}}
\caption{Angular ($\phi$) distribution of the helicity asymmetry,
$A$,
for different contributions in the $\gamma p\to\pi^0\pi^0 p$
channel for a $\gamma p$ energy of 
$\sqrt{s}=1400\textrm{ MeV}$.} 
\label{fig:results3}
\end{figure}

\begin{figure}
\centerline{\protect\hbox{
\psfig{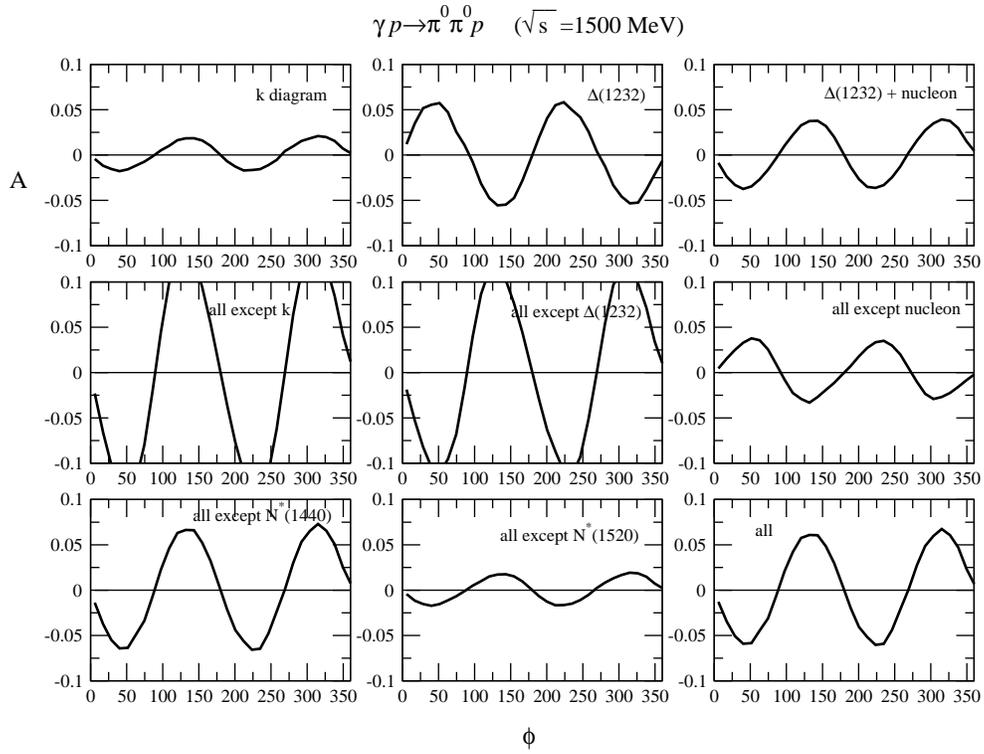}}}
\caption{Same as Fig.~\ref{fig:results3}
 for $\sqrt{s}=1500\textrm{ MeV}$.} 
\label{fig:results4}
\end{figure}

Next we show the results for the $\gamma p\to\pi^0\pi^0 p$
channel, for which there are prospects to be measured at Mainz
\cite{beck}.
In Figs.~\ref{fig:results3} and \ref{fig:results4} we show the
contribution of different mechanisms, or combinations between
them, for two different energies. 
In the $\pi^0\pi^0$ channel there are less
mechanisms allowed than in the $\pi^+\pi^-$ case,
like, for instance, the $\Delta$KR term (which gives the most
important contribution in the $\pi^+\pi^-$ channel).
From left to right and up to down the plots in 
Figs.~\ref{fig:results3} and \ref{fig:results4} represent:
$k$ diagram (which is important in the total cross section
\cite{roca}), $\Delta(1232)$ (diagrams $h$, $k$, $m$, $o$ and
$p$), $\Delta(1232)$ plus nucleon terms, all the mechanisms
except $k$, all except $\Delta(1232)$, all except nucleon, all
except $N^*(1440)$, all except $N^*(1520)$ and , finally, the
full model.

In this channel, apart from the condition
$A(\phi)=-A(2\pi-\phi)$, there is another extra condition which
is $A(\phi)=A(\phi+\pi)$. This happens since the two $\pi^0$ are
identical particles and the observables cannot depend on
permuting the two $\pi^0$, but the exchange of the two pions
means the change $\phi\to\phi+\pi$ (see
Fig.~\ref{fig:kinematics}). The conditions $A(\phi)=-A(2\pi-\phi)$
and $A(\phi)=A(\phi+\pi)$ imply that, in the series of
Eq.~(\ref{eq:series}) only the $n=$even terms are possible. That
is why the angular dependence of the helicity asymmetry for the
$\pi^0\pi^0$ channel manifests, essentially, a $\sin(2\phi)$
shape. For this reason, the shape for this channel is less rich
in variety of structures that in the $\pi^+\pi^-$ case, since
the $\sin(\phi)$, $\sin(3\phi)$,... terms are forbidden.
Nonetheless, despite the plots in
Figs.~\ref{fig:results3} and \ref{fig:results4} manifesting
mainly a $\sin(2\phi)$ dependence, one can see there
the important
role of interferences in determining the strength and phase
(sign) of the distributions, and the energy dependence of the
effect. For instance, for $\sqrt{s}=1400\textrm{ MeV}$, despite
the nucleon mechanisms giving a vanishing contribution by
themselves (not shown in the figure), the interference with the
$\Delta(1232)$ mechanisms produces a very small asymmetry. 
Another example, quite spectacular, is the role of the $k$ diagram
since, in spite that 
by itself gives a similar distribution for both energies, the
distribution removing it from the full model looks
dramatically different in the two energies: in fact, at
$1400\textrm{ MeV}$, the angular distribution of the 
asymmetry is negligible if
one removes the $k$ mechanism from the full model, 
while at $1500\textrm{ MeV}$ the effect is very large.
The $N^*(1520)$ reduces the strength at $1500\textrm{ MeV}$ 
if it is removed from the full model, while no
significant effect is visible at $1400\textrm{ MeV}$. Therefore,
in spite the $\pi^0\pi^0$ channel
 having less richness in different shapes for
the distribution, the large variation in the strength and sign
stresses the importance of this observable in elucidating the
mechanisms involved in the reaction.\\

The calculations done in the present work are just an example of
the type of studies that one can make, in
the sense that other energies, angles or kinematical cuts can be
implemented, but it serves as an example of the strong dependence
on the internal mechanisms and interferences 
that is obtained from this kind of polarization experiments.

\section{Conclusions}

We have made calculations of angular distributions of helicity
asymmetries in $\gamma p\to\pi^+\pi^- p$ and
 $\gamma p\to\pi^0\pi^0 p$ for the interaction of
circularly polarized photons with unpolarized protons. We have
used a well tested theoretical model successfully applied in the
evaluation of several unpolarized observables.
The study of polarization
observables of the kind of those discussed in the present work,
can serve to challenge the theoretical models when more 
demanding refinement of the details can be crucial.  We have
shown the strong dependence of the shape and strength of the
calculations on the internal mechanisms and interferences among
different contributions to the process. We have shown that,
in spite that 
some mechanisms do not give structure by themselves, they can
be crucial to produce the final result due to subtle
interferences with other mechanisms. Furthermore, mechanisms
which give a small contribution in unpolarized observables, like
total cross sections, can be of strong relevance in the
contribution to the difference between the polarization cross
sections. Therefore, these polarization observables have
different sensitivities to the internal details of the model than
other observables.

Further experimental results would be of  importance to
discriminate between models but being aware that sizeable
discrepancies between theoretical and experimental results can be
due to small details which are irrelevant when applying the model
to the evaluation of unpolarized observables.

\section{Acknowledgments}
I acknowledge support from the
Ministerio de Educaci\'on, Cultura y Deporte.
I thank fruitful discussions with E.~Oset and
M.~J.~Vicente~Vacas.
This work is partly supported by DGICYT contract number BFM2003-00856,
and the E.U. EURIDICE network contract no. HPRN-CT-2002-00311.


\end{document}